\documentclass[aps,prl,twocolumn,superscriptaddress,showpacs]{revtex4}
\usepackage[intlimits]{amsmath}
\usepackage{amsfonts}
\usepackage{graphics}
\usepackage{subfigure}
\usepackage[usenames]{color}
\usepackage{epstopdf,epsfig}

\usepackage{indentfirst}

\usepackage{graphicx}
\usepackage{amsmath}
\usepackage{amsfonts}
\newcommand\be            {\begin{equation}}
\newcommand\bea           {\begin{equation}\begin{array}l\displaystyle}
\newcommand\ee            {\end{equation}}
\newcommand\bes           {\begin{subequations}}
\newcommand\esu           {\end{subequations}}

\newcommand\labl[1]       {\label{#1}\ee}

\newcommand{\ud}{\mathrm d}

\newcommand\mc            {\mathcal}

\newcommand\p            {\partial}
\newcommand\psid         {\psi^{\dagger}}
\renewcommand\th           {\theta}

\def\3pt#1#2#3{{\langle{#1}\vert{#2}\vert{#3}\rangle}}

\begin{document}

\title{Infinite-time Average of Local Fields in an Integrable \\ Quantum Field Theory after a Quantum Quench}
\author{G. Mussardo}
\affiliation{SISSA and INFN, Sezione di Trieste, via Bonomea 265, I-34136, 
Trieste, Italy}
\affiliation{International Centre for Theoretical Physics (ICTP), 
I-34151, Trieste, Italy}

\begin{abstract}
\noindent 
The infinite-time average of the expectation values of local fields of any interacting quantum theory after a global quench process are key quantities for matching theoretical and experimental results. For quantum integrable field theories, we show that they can be obtained by an ensemble average that employs a particular limit of the Form Factors of local fields and quantities extracted by the Generalized Bethe Ansatz. 

\end{abstract}
\pacs{05.30.Ch, 05.30.Jp, 11.10.Gh, 11.10.Kk}
\maketitle

The aim of this paper is to set up a statistical ensemble formula for explicitly computing the infinite-time average of the Expectation Value (EV) of local fields in a $(1+1)$ dimensional Quantum Integrable Field Theory (QIFT) \cite{notedisclaimer} after a quantum quench, i.e. after an abrupt change of the parameters of the Hamiltonian.  

The subject of quantum quenches have recently attracted a lot of attention, both from experimental and theoretical point of view, see for instance \cite{Weiss,Deutsch,CC,SilvaReview,IC,Berges,Rigol,FM,CEF,Gurarie}. QIFT's are special continuum models of quantum many-body systems: they are special for the presence of an infinite number of conservation charges ${\cal Q}_n$ that strongly constrain their scattering processes and their dynamics (see, e.g. \cite{GMUSSARDO} and references therein). In a situation of out-of equilibrium dynamics, one expects that the asymptotic infinite-time regime of these theories will violate the ergodicity property and therefore its properties could not be recovered by the usual Gibbs ensemble based only on the Hamiltonian. Indeed, it has been advocated in \cite{Rigol} that to describe the stationarity properties of these integrable systems one has to consider a Generalized Gibbs Ensemble (GGE), i.e. an ensemble that not only employs the Hamiltonian but also {\em all} the other conserved charges. Such a hypothesis has be shown to be valid in a series of examples, among which those studied in \cite{FM,CEF,Gurarie}, and has acquired by now a well-established level of consensus.  


Yet, despite important advances on many topics, an explicit formula for computing the infinite-time average of the EV of local fields in QIFT has been so far elusive. This formula is put forward and proved in this paper: it concerns with the following identity 
\be 
\langle {\cal O}\rangle_{DA} \,=\,\langle {\cal O}\rangle_{GGEA} \,\,\, , 
\label{BasicIdentity}
\ee 
where the two quantities of this equation are defined hereafter.  Let ${\cal O}(x,t)$ be a local field of this QIFT and $\langle \psi_0|{\cal O}(t)| \psi_0 \rangle$ its expectation value on a macroscopic state $|\psi_0\rangle$,  not an eigenstate of the Hamiltonian. As shown in \cite{CC}, this state encodes all the information about the quench process. Being a macroscopic state, $|\psi_0\rangle$ is necessarily made of an infinite superposition of multi-particle states \cite{FM} but the statistical nature of this state is more interesting than that and will be discussed in more detail later. Let's now define the Dynamical Average (DA) of the field ${\cal O}(x,t)$ on $|\psi_0\rangle$ as the infinite time average after the quench at $t=0$
\be 
\langle {\cal O}\rangle_{DA} \,\equiv \, \lim_{t \rightarrow \infty} \frac{1}{t}\, \int_0^t \, dt \,  
\, {\langle \psi_0 |\cal O}(0,t)| \psi_0\rangle \,\,\,.
\label{DynamicalAverage}
\ee
In infinite volume, or with periodic boundary conditions on a finite interval $L$, this average is independent on $x$ by the translation invariance of the theory. 
 
The compact definition of the Generalize Gibbs Ensemble Average (GGEA) entering eq. (\ref{BasicIdentity}) is given by   
\be
\langle {\cal O}\rangle_{GGEA}\,\equiv\, 
\sum_{n=0}^\infty\frac1{n!}
\int_{-\infty}^\infty 
\left(\prod_{i=1}^n\frac{\ud\theta_i}{2\pi} f(\theta_i) 
\right) 
\3pt{\overleftarrow{\th}}{\mc O(0,0)}{\overrightarrow{\th}}_\text{conn}\,,
\label{EnsembleAverage} 
\ee
where $|\overrightarrow{\theta}\rangle \equiv |\theta_1,\dots,\theta_n\rangle$ ($\langle \overleftarrow{\theta}| \equiv \langle \theta_n,\dots,\theta_1|$) denotes the asymptotic multi-particle states of the IQFT expressed in terms of the rapidities $\theta_i$, with relativistic dispersion relation $E(\theta)  =  m \cosh\theta$ , $ p(\theta)  =  m \sinh\theta$.  

The GGEA employs the connected diagonal Form Factor (FF) of the operator ${\mc O}$, which are {\em finite} functions of the rapidities defined as 
\begin{eqnarray}
&& \3pt{\overleftarrow{\th}}{\mc O}{\overrightarrow{\th}}_\text{conn} \equiv F_{2n,{\text conn}}^{{\cal O}}(\theta_1,\ldots\,\theta_n) 
\label{connectedFF} \\
&& = FP\left(\lim_{\eta_i\to0} \3pt{0}{\mc O}{\overrightarrow{\th},\overleftarrow{\th} - i\pi+i \overleftarrow{\eta}}\right) 
\nonumber 
\end{eqnarray}
where $\overleftarrow{\eta} \equiv (\eta_n,\dots,\eta_1)$ and $FP$ in front of the expression means taking its finite part, i.e. omitting all the terms of the form $\eta_i/\eta_j$ and $1/\eta_i^p$ where $p$ is a positive integer, in taking the limit $\overleftarrow{\eta} \rightarrow 0$ in the matrix element given above. Formula (\ref{EnsembleAverage}) also employs the filling factor $f(\theta)$ of the one-particle state 
\be
f(\theta_i) \,=\, (e^{\epsilon(\theta_i)}-S(0))^{-1}
\,\,\,,
\label{fillingfactor}
\ee
where $S(\theta)$ the exact two-body S-matrix of the model while the pseudo-energy $\epsilon(\th)$ is solution of the Generalized Bethe Ansatz (GBA) equation based on all the conserved charges of the theory \cite{MC}, see eq.(\ref{GBA}) below. 

\vspace{1mm}
\noindent
It is worth making a series of comments: 

\vspace{1mm}
\noindent
(a) the final formula (\ref{EnsembleAverage}) of the Generalize Gibbs Ensemble Average may be regarded as a generalization of the so-called LeClair-Mussardo (LM) formula \cite{LM}, previously established in the context of pure thermal equilibrium. The main difference between the two's is that while the LM formula employs the Thermodynamics Bethe Ansatz \cite{YY,Zam}, the expression (\ref{EnsembleAverage}) instead employs  the Generalized Bethe Ansatz \cite{FM,MC}, i.e. the formalism that takes into account all the conserved charges of the initial state used in the quench process. 

\vspace{1mm}
\noindent
(b) the expression (\ref{EnsembleAverage}) consists of a well-defined and, usually, fast convergent series \cite{note3}. Particularly useful is the fast convergence of the series, because it permits to compute the infinite-time averaged EV, for all intents and purposes, by employing just the first few terms, saving then a lot of analytic and numerical efforts. 

\vspace{1mm}
\noindent
(c) It is also interesting to mention that, restoring in the IQFT a $\hbar$ dependence, the limit $\hbar\rightarrow 0$ of the formula 
(\ref{EnsembleAverage}) solves a long-standing problem of purely mathematical physics, i.e. how to determine the infinite time-averages in purely {\em classical} relativistic integrable models when one is in presence of the so-called {\em infinite-gap} solutions \cite{MDD}. 

\vspace{1mm}
Before embarking in the proof of the identity (\ref{BasicIdentity}), it is useful to spell out its content by means of the simplest QIFT, i.e. the free theory. Although elementary, the important pedagogical value of this example is to show clearly the necessity to employs in the identity (\ref{BasicIdentity}) the Generalized Gibbs Ensemble Average (\ref{EnsembleAverage})  \cite{note4}. 


In the infinite volume, the solution of the free eq. of motion  
$
(\Box + m^2) \,\phi(x,t) \,=\, 0 \,
$ 
is 
\be
\phi(x,t) = \int_{-\infty}^{\infty} \frac{d\theta}{2 \pi}
\left[
A(\theta) \,e^{-i (E(\theta) t-p(\theta) x)} + c.c. 
\right] 
\label{exactsolutionfreetheory}
\ee
with $[A(\theta), A^{\dagger}(\theta')] = 2 \pi \delta(\theta-\theta')$. Such a dynamics is supported by the infinite number of {\em non-local} conserved quantities given by all the mode number occupations 
\be
N(\theta) = \frac{1}{2 \pi} |A(\theta)|^2 \,\,\,\,\,, \,\,\,\, \forall \theta \,\,\,.
\label{modeoccupation}
\ee
As shown in the Supplementary Material, one can can also find the infinite set of {\em local} conserved charges 
${\cal Q}_n$ made of two sets ${\cal Q}_n^+$ and ${\cal Q}_n^-$, the first even under the $Z_2$ space-parity, the second odd. 
${\cal Q}_0^+$ and ${\cal Q}_0^-$, are respectively the energy and momentum of the field, while the others, up to normalization, can be written 
as      
\begin{eqnarray}
{\mathcal Q}_{n}^{\pm} & \,= \, &  
m^{2n+1} \,\int \frac{d\theta}{2\pi} \,|A(\theta)|^2 \, q_n^{\pm}(\theta) \,\,\,\label{localconservationlaws} \,\,\,,
\end{eqnarray} 
where
\be
\begin{array}{c}
q_n^{+}(\theta)=\cosh[(2n+1)\theta] \,\,\,,\\
q_n^-(\theta) =\sinh[(2n+1)\theta] \,\,\,.
\end{array}
\label{q+-}
\ee 
The multi-particle states are common eigenvectors of all these conserved quantities, with eigenvalues 
\be 
{\mathcal Q}_{n}^{\pm} \,| \theta_1,\ldots,\theta_k \rangle \,=\, m^{2n+1} \left(\sum_{i=1}^k \,q_n^{\pm}(\theta_i)\right) \, 
| \theta_1,\ldots,\theta_k \rangle  \,\,\,.
\label{actionconservedcharges}
\ee
It is pretty evident that the knowledge of the mode occupation $|A(\theta)|^2$ fixes all the local charges but, under general mathematical and physical conditions, it is also true the viceversa, alias that the ${\mathcal Q}_{n}^{\pm}$'s fix the $|A(\theta)|^2$. Being linearly related one to the other, the two types of conservation laws are then essentially interchangeable.

It must be stressed that eqs. (\ref{modeoccupation}), (\ref{localconservationlaws}) and (\ref{actionconservedcharges}) holds exactly the same also in interacting QIFT (as the ShG model, for instance), where the only thing to do is to substitute, in eqs.(\ref{modeoccupation}) and (\ref{localconservationlaws}), $|A(\theta)|^2 \rightarrow |Z(\theta)|^2$, where $Z(\theta)$ and $Z^{\dagger}(\theta)$ satisfy the Faddev-Zamolodchikov algebra involving the exact $S$-matrix
\[
Z(\theta_1) Z^{\dagger}(\theta_2) \,=\,S(\theta_1-\theta_2) Z^\dagger(\theta_2) Z(\theta_1) + 2\pi\,\delta(\theta_1-\theta_2) \,\,\,.
\label{FZalgebras}
\]

In the free theory, the exact solution (\ref{exactsolutionfreetheory}) of the eq. of motion allows us to easily compute the DA of any local function $F[\phi(x,t)]$, defined with a proper normal-ordering of the operators. Since we are interested in field configurations with {\em finite energy density}, our theory has to be defined on a circle of length $L$ and then send $L\rightarrow \infty$ so that $\hat E=E/L$ is always finite, even in this limit. The momenta of the particles will be quantized in unit of $2 \pi/L$ which become dense when $L\rightarrow \infty$. The initial state $|\psi_0\rangle$ fixes the modes $A(\theta)$ and $A^\dagger(\theta)$ through the condition $<\psi_0 |\phi(x,0)|\psi_0> = \psi_0(x)$, where $\psi_0(x)$ is a real periodic function $\psi_0(x) = \psi_0(x+L)$, such that $\frac{1}{2 L} \int_0^L [(\partial_x \psi_0)^2 + m^2 \psi_0^2] dx = \hat E$.  

\begin{figure}[t]
\centering
$
\begin{array}{cc}
\includegraphics[width=0.4\textwidth]{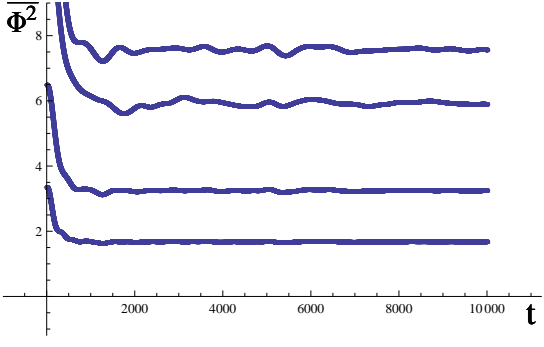} &
\end{array}
$
\caption{$\overline \Phi^2(t) \equiv \frac{1}{t}\, \int_0^t \, dt \,  
\, \langle \psi_0 |\phi^2(t)| \psi_0\rangle$ as a function of the time $t$, for different initial states $|\psi_0\rangle$ with the {\em same} EV of the energy density. The DA, as defined in eq.(\ref{DynamicalAverage}), are the asymptotic values of these curves. }
\label{spread}
\end{figure}

Let's now consider in the free theory a series of quenches, whose initial states have in common only the same energy density $E/L$. Since the energy is a very degenerate observable, each of these quenches corresponds to initial states having different EV of all other conserved charges. The typical outputs for an observable as $:\phi^2:$ are shown in Figure \ref{spread}: the strong dependence on the initial data is evident from the large spread of these DA. These features are easily explained. Focus the attention on the DA of this infinite set of  operators (easily computable by a phase-stationary argument)  
\begin{eqnarray} 
&& \langle :\phi^2: \rangle_{DA} \,=\, \int \frac{d\theta}{2\pi}\, |A(\theta)|^2\,\equiv\, b \,\,\,, \label{vevphin}\\ 
&& \langle :\phi^{2n}: \rangle_{DA} \,=\, (2 n -1) !! \, b^n \label{vevphinn}\,\,\,.
\end{eqnarray}
Since they explicitly depend on the initial condition through the $|A(\theta)|^2$'s,  these DA can {\em never} collapse on a single value or be derived by a Gibbs Ensemble average involving only the Hamiltonian  
\be 
\langle :\phi^{2n}: \rangle_{DA} \,\neq \, Z^{-1} \,\text{Tr}\, ( :\phi^{2n}: \, e^{-\beta H}) \,, 
\ee
even if all initial states share the same energy. 

Notice, however, that the DA (\ref{vevphin}) and (\ref{vevphinn}) can be put precisely in the form of the Generalized Gibbs Ensemble Average 
(\ref{EnsembleAverage}), since the pseudo-energy and the non-zero connected FF of these operators are in this case
\begin{eqnarray*}
&&\epsilon(\theta) = \log (1+ |A(\beta)|^{-2}) \,\,\,,\\
&& \langle{\overleftarrow{\theta_m}}|:\phi^{2n}(0):|{\overrightarrow{\theta_m}}\rangle_\text{conn}\,=\,2^n (2 n)! \delta_{n,m}\,\,\,. 
\end{eqnarray*}
(see the Supplementary Material). In summary,  we have verified that the identity (\ref{BasicIdentity}) holds in the integrable 
free theory. Moreover, all the DA of free theory can be recovered expanding in $\alpha$ the GGEA generating function
\be 
\langle \exp[ i \alpha \phi]\rangle_{GGEA} \,=\, \exp \left[-\frac{\alpha^2}{2} \int \frac{d\theta}{2\pi} |A(\theta)|^2  
\right] \,\,\,.  
\label{stationargeneratingfunction}
\ee

\vspace{1mm}

Let's now proceed to the general proof of the identity (\ref{BasicIdentity}) for an interacting QIFT, with a two-body $S$-matrix $S(\theta)$ assumed to be of fermionic type, i.e. $S(0)=-1$.  Expanding the initial state $|\psi_0\rangle$ on the basis of the multi-particle states, which are common eigenvectors of $H$ and all higher charges, its general form is  
\be
|\psi_0\rangle = \sum_{n=0}^{\infty} \frac{1}{n!} \int \prod_{i=1}^n\left(\frac{d\theta_i}{2\pi}\right)\, K_n(\theta_1,\ldots,\theta_n)\, 
|\theta_1,\ldots,\theta_n\rangle\,\,\,.\label{multiparticleexpansion}
\ee
Posing  $E_n = m \sum_{i=1}^n \cosh\theta_{i}$, at any later time $t$ the EV of a local observable ${\cal O}(x,t)$ on $|\psi_0\rangle$ is  
\begin{eqnarray*}
&&
\frac{\langle \psi_0| {\cal O}(t) | \psi_0\rangle}{\langle \psi_0 | \psi_0\rangle} =Z^{-1} \sum_{m,n=0}^\infty \frac{1}{n!m!} 
\int 
\prod_{i=1}^n \prod_{j=1}^m 
\left(
\frac{d\theta_i}{2\pi} 
\frac{d\theta_j'}{2\pi}
\right)
\\
&& e^{-i t (E_m-E_n)} 
K_m^*(\{\theta' \}) K_{n}(\{\theta\}) \langle \theta_{m}'\ldots \theta_{1}' | {\cal O}(0)| |\theta_{1}\ldots \theta_{n}\rangle
\end{eqnarray*}
($Z=\langle \psi_0 | \psi_0\rangle$). Taking the Dynamical Average, one inevitably ends up to the so-called 
Diagonal Ensemble 
\begin{eqnarray}
 \langle {\cal O} \rangle_{DA} &\,= &\,Z^{-1} \sum_{n=0} \frac{1}{n!} \int \prod_{i=1}^n
\left(
\frac{d\theta_i}{2\pi}\right) |K_n(\theta_1,\ldots,\theta_n|^2 \nonumber \\
& & \,\,\,\,\times  \langle \theta_{n}\ldots \theta_{1} | {\cal O}(0)| |\theta_{1} 
\ldots \theta_{n}\rangle\, \,\,\,.\label{DiagonalEnsemble}
\end{eqnarray}
As it stands, however, this expression is highly problematic: all terms of the sum in the numerator as well as those present in $Z$ are in fact {\em divergent}. In the latter, the divergencies come from the normalization of the eigenstates, $\langle \theta'_{m},\ldots \theta'_{1} | \theta_{1}\ldots \theta_{n}\rangle = \prod_i \delta(\theta'-\theta)$, which gives rise to $[\delta(0)]^n$ when $\theta'_i=\theta$. In the former, the divergencies come from the Form Factor $\langle \theta'_{m}\ldots \theta'_{1} | {\cal O}(0)| |\theta_{1}\ldots \theta_{n}\rangle $, once evaluated at $\theta'_i = \theta_i$. These divergencies are an unavoidable consequence of the kinematical pole structure of the Form Factors \cite{Smirnov}. 

The first cure of the divergencies is to define the theory on a finite interval $L$. In this case, for large but finite $L$, the rapidities of the $n$-particle states entering (\ref{multiparticleexpansion}) are solutions of the Bethe Equation 
\be
 m L\sinh\theta_i +  \sum_{k\neq i} \varsigma(\theta_i-\theta_k) \,=\, \frac{2\pi N_i}{L} \,\,\,,\,\,\,
i=1,\ldots,n 
\label{BETHE}
\ee
where $\varsigma(\theta) = - i \log S(\theta)$ is the phase-shift and $\{N_i\}$ is a sequence of increasing integers. Let's denote the finite volume eigenstates associated to the integers $\{N_i\}$ as $| \theta_1,\ldots,\theta_n\rangle_L$: the corresponding density of states is given by the Jacobian $J(\theta_1,\ldots,\theta_n)= {\rm det}\, { J}_{jk} $, with ${J}_{jk} \,=\,\frac{\partial {\cal J}_j}{\partial \theta_k}$. The functions ${\cal J}_i$ are given by the derivative of the r.h.s. of (\ref{BETHE}) 
\be
{\cal J}_i \,(\theta_1,\ldots,\theta_n) =\, m L \cosh\theta_i + \sum_{k\neq i} \varphi(\theta_i-\theta_k)  \,\,\,,
\label{densitystate}
\ee
with the kernel $\varphi(\theta) = - i \frac{d}{d\theta} \log S(\theta)$. 

There is now a relation between the diagonal Form Factors in finite volume and the infinite-volume connected Form Factors 
defined in eq.(\ref{connectedFF}) \cite{PozsayTakacs}  
\begin{eqnarray}
\langle \theta_1,\dots,\theta_n | {\mc O}| \theta_1,\dots,\theta_n\rangle_\text{L} \,=\,\frac{1}{{\cal J}_n(\theta_1,\ldots,\theta_n)} \nonumber \\ 
\times 
\sum_{\{\theta_+\} \bigcup \{\theta_-\}} F_{2l,{\rm conn}}^{{\cal O}}(\theta_-) \, \overline{\cal J}_{n-l}(\{\theta_+\},\{\theta_-\}) \,\,\,,
\label{finitevsinfinitediagonalFF}
\end{eqnarray}
where the sum runs on all possible bipartite partitions of the set of rapidities $\{\theta_1,\ldots,\theta_n\}$ in two disjoint sets made by $l$ and $n-l$ rapidities, and $\overline{\cal J}_{n-l}(\{\theta_+\},\{\theta_-\}) = {\rm det}\, J_+$ is the restricted determinant of the sub-matrix ${\cal J}_+$ corresponding to the particles in the set $\{\theta_+\}$ in the presence of those in $\{\theta_-\}$. Notice that the relation (\ref{finitevsinfinitediagonalFF}) involves the kernel $\varphi(\theta)$ of the Bethe Ansatz eqs. (\ref{densitystate}), as explicitly shown in the Supplementary Material. 

Let's now focus the attention on the initial state: if the state $|\psi_0\rangle$ is statistically characterized by the EV of all its conserved charges $\langle\psi_0 |{\cal Q}_n^\pm|\psi_0\rangle = L \overline {{\cal Q}_n^\pm}$, it can be shown (see the Supplementary Material) 
that the quantities $|K_n(\theta_1,\ldots,\theta_n)|^2$ factorize in terms of a function $|K(\theta)|^2$ 
\be
|K_n(\theta_1,\ldots,\theta_n)|^2 \,=\,\prod_{i=1}^n |K(\theta_i)|^2 \,\,\,,
\ee
and moreover $K(\theta)$ can be always expressed in terms of an infinite set of variables $\{\alpha_n^{\pm}\}$, conjugated to the conserved charges ${\cal Q}_n^\pm$ as 
\begin{eqnarray}
&& |K(\theta)|^2 \,=\,e^{-\epsilon_0(\theta)} \,\,\,,\\ 
&& \epsilon_0(\theta) \,=\,\sum_{n=0}^\infty(\alpha_n^{+} q_n^+(\theta) + \alpha_n^{-} q_n^-(\theta)) \,\,\,,\nonumber 
\end{eqnarray}
with the functions $q_n^{\pm}(\theta)$ given in (\ref{q+-}). 

With all the information collected above, let's now come back to the Dynamical Average (\ref{DiagonalEnsemble}):
with the regularization given by the finite interval $L$, its r.h.s. can be written as  
\be
\langle {\cal O} \rangle_{DA} \,=\,\lim_{L\rightarrow\infty} \frac{{\rm Tr} \, \left(e^{- {\cal H}} {\cal O}\right)_L}{\left({\rm Tr}\, e^{- {\cal H}}\right)_L}
\,\,\,,
\label{finitevolumeregtr}
\ee
where ${\cal H}$ is the {\em generalized} Hamiltonian that includes all the conserved charges
\[
{\cal H} \,=\,\sum_{n=0}^{\infty} \left[\alpha_n^{+} {\cal Q}_n^+ + \alpha_n^{-} {\cal Q}_n^-\right] \,\,\,. 
\]
One can now easily repeat the argument given in \cite{LM,PozsayTakacs} and show that the r.h.s. precisely coincides with the Generalized Gibbs Ensemble average (\ref{EnsembleAverage}), where the function $\epsilon(\theta)$ satisfies the non-linear integral equation of the Generalized Bethe Ansatz \cite{MC}
\begin{eqnarray}
\epsilon(\theta) & \,=\, &
\sum_{n=0}^{\infty} \left[\alpha_n^{+} q^+(\theta) + \alpha_n^{-} q^-(\theta) \right] \nonumber\\
&-& \int\frac{d\theta'}{2\pi} \varphi(\theta-\theta') \,\log\left(1+e^{-\epsilon(\theta')}\right) \,\,\,.
\label{GBA}
\end{eqnarray}
This concludes the general proof of the identity (\ref{BasicIdentity}). A pragmatic approach to find the functions $\epsilon(\theta)$ and 
$\epsilon_0(\theta)$ given the initial state $|\psi_0\rangle$ has been recently proposed in \cite{CK}.  Applications to quench processes in the Sinh-Gordon model (both at the quantum and classical level) are presented in \cite{MDD}. It must be stressed that the same formalism can be applied to compute infinite-time average of local fields in the Lieb-Liniger model, a system which recently attracts a lot of interest for the on-going experiments in cold-atom physics: indeed, as shown in \cite{KMT}, to recover the Lieb-Liniger results one can take advantage of the fact that the Lieb-Liniger model may be reached by taking the non-relativistic limit of the Sinh-Gordon model, whose Form Factors are all known. An example relative to quench processes in the Lieb-Liniger system is presented in the Supplementary Materials.


\vspace{0.1cm}

\noindent
{\em Acknowledgements:} I would like to thank P. Assis and in particular A. De Luca for discussions. 
This work is supported by the IRSES grants QICFT.

\vspace{0.1cm}
\noindent
{\em Note Added}. Recently there has been another proposal \cite{CE} to compute the EV, purely based on the Bethe Ansatz and 
checked for the free case of the quantum Ising model. Although very similar to the one presented here, it remains to see how  
it applies to interactive case. 
.

\newpage

\clearpage 
\setcounter{equation}{0}
\renewcommand{\theequation}{S\arabic{equation}}
\setcounter{page}{1}

\onecolumngrid

\begin{center}
{\Large{\bf Supplementary Material}}
\end{center}

\section{ Local conserved charges} 

To find the infinite set of {\em local} conserved quantities in the free theory, it is convenient to go in the light-cone coordinates 
$\tau = t+x $ and $\sigma= x-t$, where the eq. of motion becomes $\phi_{\sigma\tau} \,=\,m^2 \,\phi $ and, as a consequence, 
there is the infinite chain of conservation laws 
\be
\begin{array}{c}
\partial_\tau  \phi^2_{n \sigma} \,=\,  m^2 \,\partial_\sigma  \phi^2_{(n-1) \sigma} \,\,\, \\
\partial_\sigma \phi^2_{n \tau} \,=\,  m^2 \,\partial_\tau \phi^2_{(n-1) \tau} \,\,\,,
\end{array}
\ee
($n = 1,2, \cdots $), 
where $\phi_{n \sigma} = \partial^n_\sigma \phi$ and analogously for $\phi_{n \tau}$. These equations are of the 
general form 
$$
\partial_{\tau} A = \partial_{\sigma} B $$ 
and, going back to the coordinates $(x, t)$, they 
become the continuity equation 
$$
\partial_t (A+B) \,=\,\partial_x (B-A)\,\,\,,
$$
so that the associate conserved charges are $Q = \int dx (A+B)$.

For the free-theory we have then the following set of conserved charges 
\begin{eqnarray}
& Q_n & \,= \,  \int dx \left[\frac{1}{2} \phi^2_{(n+1) \sigma} + \frac{m^2}{2} \phi^2_{n \sigma} \right] \,\,\,\\
& Q_{-n} & \,= \,  \int dx \left[\frac{1}{2} \phi^2_{(n+1) \tau} + \frac{m^2}{2} \phi^2_{n \tau} \right] \,\,\,\nonumber 
\end{eqnarray}
Taking the sum and the difference of these quantities, we can define the even and odd conserved charges 
\begin{eqnarray}
{\mathcal Q}_n^+ & \,= \, & (Q_n + Q_{-n}) = 
\frac{1}{2} \int dx \left[ \phi^2_{(n+1) \sigma} + \phi^2_{(n+1) \tau} + m^2 (\phi^2_{n \sigma} + \phi^2_{n \tau}) \right] \,\,\,\\
{\mathcal Q}_n^- & \,= \, &  (Q_n - Q_{-n}) = 
\frac{1}{2} \int dx \left[ \phi^2_{(n+1) \sigma} - \phi^2_{(n+1) \tau} + m^2 (\phi^2_{n \sigma} - \phi^2_{n) \tau}) \right] \,\,\,\nonumber
\end{eqnarray} 
It is now easy to see that they can be expressed in terms of the mode occupation of the field: using the expansion
(\ref{exactsolutionfreetheory}), we have  
\begin{eqnarray}
{\mathcal Q}_n^+ & \,= \, &  
m^{2n+1} \,\int \frac{d\theta}{2\pi} \,|A(\theta)|^2 \, \cosh[(2 n+1) \theta] \,\,\,\label{evenQ}\\
{\mathcal Q}_n^- & \,= \, & m^{2n+1} \,\int \frac{d\theta}{2\pi} \, |A(\theta)|^2 \, \sinh[(2 n+1) \theta] 
 \,\,\,\label{oddQ}
\end{eqnarray} 
${\cal Q}_0^+$ and ${\cal Q}_0^-$ correspond respectively to the energy and the momentum of the field. In the quantum field theory 
interpretation, the equations above imply that each particle state $| \theta \rangle$ of rapidity $\theta$ is a common eigenvectors of all these conserved quantities, with eigenvalues 
\be 
{\mathcal Q}_n^+ \,| \theta \rangle \,=\, m^{2n+1} \,\cosh[(2 n +1) \theta] \, | \theta \rangle    
\,\,\,\,\,\,\,\,\,\,\,
,
\,\,\,\,\,\,\,\,\,\,\,
{\mathcal Q}_n^- \,| \theta \rangle \,=\, m^{2n+1} \,\sinh[(2 n +1) \theta] \, | \theta \rangle  \,\,\,.
\ee

For an interactive integrable model as the Sinh-Gordon, using the light-cone coordinates and a mapping to the KdV equation, one can also recover the infinite set of conserved charges also for this model (details can be found in \cite{MDD}). In this interacting 
model one can uses the Inverse Scattering Transform and introduce the operators 
$Z(\theta)$ and $Z^\dagger(\theta)$ (the analogous of $A(\theta)$ and $A^\dagger(\theta)$ of the free theory) 
which satisfy the Faddev-Zamolodchikov algebra  
\be 
Z(\theta_1) Z^\dagger(\theta_1) \,=\,S(\theta_1-\theta_2) \, Z^\dagger(\theta_1) Z^\dagger(\theta_2) + 2\pi \delta(\theta_1-\theta_2) 
\,\,\,,
\label{FZalgebra}
\ee
where $S(\theta)$ is the exact 2-body scattering matrix. The conserved charges assume then the same form of (\ref{evenQ}) and 
(\ref{oddQ}) just using the substitution 
$$
|A(\beta)|^2 \rightarrow |Z(\beta)|^2 \,\,\,.
$$

\section{Generalized Gibbs Ensemble Averages in free theory} 
In the free theory the simplest way to derive the GGE density matrix is to use as infinite set of conserved quantities the mode occupations $|A(\theta)|^2$ and set  
\be
\rho_{GGE} \,=\,Z^{-1} \, \exp\left[ -\int \frac{d\theta}{2\pi} \epsilon(\theta) \, |A(\theta)|^2  \right] \,\,\,,
\label{rhofreeGGE}
\ee
The function $\epsilon(\theta)$, that plays the role of an infinite set of lagrangian multipliers, is fixed by the initial occupation numbers 
as 
$
\langle \psi_0||A(\theta)|^2|\psi_0\rangle  = (e^{\epsilon(\theta)} -1)^{-1} 
$  
\cite{CC}. 
The statistical weight given by the density matrix (\ref{rhofreeGGE}) is equivalent to  
\be
\begin{array}{l} 
\langle A(\theta) A^{\dagger}(\theta') \rangle_{GGEA} \,=\,2\pi \delta(\theta-\theta') \,\,\,,\\ 
\langle A(\theta) A(\theta')\rangle_{GGEA} =  \langle A^{\dagger}(\theta) A^{\dagger}(\theta')\rangle_{GGEA} = 0 \,\,\,,
\end{array}
\ee 
from which one can easily get the generating function 
\be 
\langle \exp[ i \alpha \phi]\rangle_{GGEA} \,=\, \exp \left[-\frac{\alpha^2}{2} \int \frac{d\theta}{2\pi} |A(\theta)|^2  
\right] \,\,\,. 
\label{stationargeneratingfunction2}
\ee
Expanding in power series in $\alpha$ the left/right hand sides and comparing equal powers in $\alpha$, one recovers the previous results (\ref{vevphin}) and (\ref{vevphinn}). 

\section{Finite-volume diagonal matrix elements  and connected Form Factors}
Here we give the first few examples of the relation that links the finite-volume diagonal matrix elements and the infinite-volume  
connected Form Factors defined in (\ref{connectedFF}). First of all, the relation (\ref{finitevsinfinitediagonalFF}) can be equivalently written as \cite{PozsayTakacs}
\be
\langle \theta_1,\dots,\theta_n | {\mc O}| \theta_1,\dots,\theta_n\rangle_\text{L} \,=\,\frac{1}{{\cal J}_n(\theta_1,\ldots,\theta_n)} \,
\sum_{\{\theta_+\} \bigcup \{\theta_-\}} F_{2l,{\rm sym}}(\theta_-) \, {\cal J}_{n-l}(\theta_+) \,\,\,,
\label{finitevsinfinitediagonalFF1}
\ee 
where, as before, the sum runs on all possible bipartite partitions of the set of rapidities $\{\theta_1,\ldots,\theta_n\}$ in two disjoint sets made by $l$ and $n-l$ rapidities, while  $F_{2l,{\rm sym}}(\theta_1,\ldots, \theta_l)$ are finite functions defined by the {\em symmetric limit}  
\begin{eqnarray}
&& F_{2n,{\rm sym}}(\theta_1,\ldots,\theta_n) \,=\, \langle \theta_1,\dots,\theta_n | {\mc O}| \theta_1,\dots,\theta_n\rangle_\text{sym} \label{eq:symmdef}\\
&& = \left(\lim_{\eta\to0} \3pt{0}{\mc
  O}{\theta_1 + i \pi + i \eta, \dots, \theta_n +i \pi + i\eta, \theta_1,\dots,\theta_n}\right)\,.\nonumber 
\end{eqnarray}
Notice that, while eq.(\ref{finitevsinfinitediagonalFF}) employs $\overline{\cal J}_{n-l}(\{\theta_+\},\{\theta_-\})$ that contains information 
both on $\{\theta_+\}$ and its complementary set $\{\theta_-\}$, eq.(\ref{finitevsinfinitediagonalFF1}) instead employs the density of states ${\cal J}_{n-l}(\theta_+)$. 

Since for any local operator ${\cal O}$ its $F_2^{\cal O}(\theta)$ is a constant, $F_2 = \langle \theta|{\cal O}(0,0)|\theta\rangle$ and therefore  $F_2 =F_{2, {\rm conn}} = F_{2, {\rm sym}}$. Expressing now $F_{2n,{\rm sym}}$ in terms of $F_{2n,{\rm conn}}$, for the next few cases we have 
\begin{eqnarray*}
&& 
F_{4, {\rm sym}}(\theta_1,\theta_2) \,=\,F_{4, {\rm conn}}(\theta_1,\theta_2) + 2 \varphi(\theta_1-\theta_2) \,F_{2, {\rm conn}} 
\,\,\,;\\ [5pt]
&& 
F_{6, {\rm sym}}(\theta_1,\theta_2,\theta_3) \,=\,F_{6, {\rm conn}}(\theta_1,\theta_2,\theta_3) + 
\left[F_{4, {\rm conn}}(\theta_1,\theta_2) \left(\varphi(\theta_1-\theta_3) + \varphi(\theta_2-\theta_3)\right) + {\rm permutations}\right] \\
&& 
+ 3 F_{2, {\rm conn}} \left[\varphi(\theta_1-\theta_2) \,\varphi(\theta_1-\theta_3) + {\rm permutations} \right] 
\nonumber \,\,\,.\\
\end{eqnarray*}
It is then clear that the finite-volume diagonal matrix elements (\ref{finitevsinfinitediagonalFF1}) 
are expressed in terms of the infinite-volume connected Form Factors $F_{2l,{\rm conn}}(\theta_1,\ldots,\theta_l)$  and the kernel $\varphi(\theta)$ of the Bethe Ansatz equations (\ref{densitystate}). 

\section{Statistical properties of the initial state}

Given the expansion (\ref{multiparticleexpansion}) of the initial state $|\psi_0\rangle$, the quantity that enters 
the infinite-time averages is its diagonal density matrix defined by 
\be
\rho_d \,=\,\sum_{n=0}^\infty \frac{1}{n!}  \int \prod_{i=1}^n
\left(\frac{d\theta_i}{2\pi}\right)|K_n(\theta_1,\ldots,\theta_n)|^2 
 |\theta_1,\ldots,\theta_n\rangle \,\langle \theta_n,\ldots,\theta_1 | \,\,\,, 
\label{diagonalK}
\ee
where 
\[
|\theta_1,\ldots,\theta_n\rangle \,=\,Z^{\dagger}(\theta_1) Z^{\dagger}(\theta_2) \cdots Z^{\dagger}(\theta_n) | 0 \rangle\,\,\,.
\]
Notice that $|K_n(\theta_1,\ldots,\theta_n)|^2$ can be assumed to be completely symmetric functions with respect any 
permutation of their variables, since for the original functions holds 
\be
K_n(\theta_1,\ldots,\theta_l,\theta_{l+1}\,\ldots \theta_n) \,=\, S(\theta_l-\theta_{l+1}) K_n(\theta_1,\ldots,\theta_{l+1},\theta_{l}\,\ldots \theta_n) \,\,\,.
\ee
and therefore
\be 
|K_n(\theta_1,\ldots,\theta_l,\theta_{l+1}\,\ldots \theta_n)|^2 \,=\, | K_n(\theta_1,\ldots,\theta_{l+1},\theta_{l}\,\ldots \theta_n)|^2 \,\,\,,
\ee
since the $S$-matrix is a pure phase. 

Let's see what kind of constraints we have on the functions $|K_n(\theta_1,\ldots,\theta_n)|^2$ 
assuming that we known all the expectation values $\overline{ {\cal Q}_n^{\pm}}$ of the conserved 
charges ${\cal Q}_n^{\pm}$ on the initial state $|\psi_0 \rangle$
\be
\langle \psi_0 | {\cal Q}_n^{\pm} | \psi_0\rangle  \equiv L \overline{ {\cal Q}_n^{\pm}} \,\,\,. 
\label{VEQn}
\ee
To this aim, consider the infinite set of partition functions  
\be
{\cal Z}_n^{\pm} \,=\, {\rm Tr}\,\left(\rho_d \, e^{-\alpha_n {\cal Q}_n^{\pm}} \right) \,\,\,,
\label{partitionfunctionn}
\ee
($n=0,1,2,\ldots$) associated to each conserved charges. For large $L$, using the extensivity property of the free energy, 
we have 
\be
{\cal Z}_n^{\pm} \,=\,e^{ -L u_n^{\pm} } \,\,\,,   
\label{largeLZn}
\ee
where $u_n^{\pm}$ is the corresponding the free energy per unit length.  
Expanding this expression, we then have 
\be 
{\cal Z}_n^{\pm} \,=\,1 - L u_n^\pm  + \frac{1}{2!} L^2 (u_n^\pm )^2 + \cdots + 
(-1)^k \frac{1}{k!} L^k (u_n^\pm )^k + \cdots
\label{expansionZn}
\ee
On the other hand, we can compute ${\cal Z}_n$ using directly its definition (\ref{partitionfunctionn})
\be
{\cal Z}_n^\pm \,=\,1 + {\cal Z}_{n,1}^\pm + {\cal Z}_{n,2}^\pm + \cdots {\cal Z}_{n,k}^\pm + \cdots
\label{explicitexpansionZn}
\ee
where ${\cal Z}_{n,k}^\pm$ is the contribution coming from the $k$-multiparticle state. To compute these terms, one needs 
though to regularize on the finite volume $L$ the square of the $\delta$-functions which comes from the scalar product of the multi-particle states. This can be done as in \cite{LM}, with the substitution 
\be
[\delta(\theta-\theta')]^2  \rightarrow  \frac{L}{2\pi} \cosh\theta \,\delta(\theta-\theta') \,\,\,.
\label{regsquaredelta}
\ee
Since the multi-particle states are eigenvectors of the conserved charges 
\be 
{\cal Q}_n^{\pm} |\theta_1,\ldots,\theta_n\rangle \,=\,\left(\sum_{i=1}^n q^{\pm}_n(\theta_i)\right)   |\theta_1,\ldots,\theta_n\rangle\,\,\,,
\ee
using the Faddev-Zamolodchikov algebra (\ref{FZalgebra}) and the regularization (\ref{regsquaredelta}), we can trace back the 
$L$ dependence in the terms ${\cal Z}_{n,k}^\pm$. To express them in a compact way 
it is useful to define the quantities 
\be 
I_{p,m}^{(n)} \,\equiv \, \int \frac{d\theta_m}{2\pi} \cosh\theta_m e^{- (p-m+1) \alpha_n  q_n^\pm(\theta_m)} \,\int
\prod_{i=1}^{m-1} \left(\frac{d\theta_i}{2\pi} \cosh\theta_i e^{-\alpha_n q_n^\pm(\theta_i)}\right) 
\, 
|K_p(\theta_1,\ldots,\theta_{m-1},\underset{p-m+1}{\underbrace{\theta_m,\theta_m,\ldots,\theta_m}})|^2\,\,\,.
\label{definitionIp}
\ee   
Then, for the first few we have 
\begin{eqnarray}
&& {\cal Z}_{n,1}^\pm = L I_{1,1}^{(n)}  \nonumber \\
&& {\cal Z}_{n,2}^\pm =  \frac{ L^2}{2}  I_{2,2}^{(n)} -  \frac{L}{2}  I_{2,1}^{(n)}  \label{singleterms}\\
&& {\cal Z}_{n,3}^\pm = \frac{L^3}{3!}  I_{3,3}^{(n)} - \frac{L^2}{2} I_{3,2}^{(n)} 
+ \frac{L}{3} I_{3,1}^{(n)}\nonumber
 \end{eqnarray}
It is easy to see that, in the quantities ${\cal Z}_{n,p}$,  the terms proportional to $L$  come from the integrals $I_{p,1}^{(n)}$, while those proportional to $ L^p$ come from $I_{p,p}^{(n)}$, with combinatorial factors $1/p$ and $1/p!$ respectively. 

If the partition function ({\ref{partitionfunctionn}) has to exponentiate for large $L$ as in (\ref{largeLZn}), the integral 
$I_{k,k}$ must scale for $k \gg 1$ as a power law in terms of some constant $\lambda_n$ 
\begin{equation}
I_{k,k} = \int \prod_{i=1}^k \left(\frac{d\theta_i}{2\pi} \cosh\theta_i e^{-\alpha_n q_n^{\pm}(\theta_i)}\right) 
|K_n(\theta_1,\ldots,\theta_n)|^2 
\,\simeq \, 
\lambda_n^k \,\,\, , 
\end{equation}
Since this power-law behavior for $I_{k,k}$ must hold {\em all} charges ${\cal Q}_n$, i.e. no matter how we vary the eigenvalue functions $q_n^\pm(\theta)$ in eq.(\ref{finalgame}), $|K_n(\theta_1,\ldots,\theta_n)|^2$ must factorize in terms of a function $K(\theta)$ as 
\be
|K_n(\theta_1,\ldots\,\theta_n)|^2 \,=\,\prod_{i=1}^n |K(\theta_i)|^2 \,\,\,.
\ee
Notice that this factorization condition only holds for the modulus square of the amplitudes $K_n(\theta_1,\ldots,\theta_n)$ and 
not for the amplitudes themselves. 

If we now assume that the partition function can be {\em exactly} expressed as in eq.(\ref{largeLZn}), and not only for large value 
of $L$, we have even a stronger result, namely that the function $|K(\theta)|^2$ coincides with $K_1(\theta)|^2$. In fact, collecting all terms proportional to $L$ in the ${\cal Z}_{n,k}^\pm$ and making their sum we have 
\be 
g_1^{(n)} \,\equiv \,I_{1,1}^{(n)} -\frac{1}{2} I_{2,1}^{(n)} + \frac{1}{3} I_{3,1}^{(n)} + \cdots (-1)^{p+1} \frac{1}{p} I_{p,1}^{(n)} + \cdots 
\ee
With the partition function ({\ref{partitionfunctionn}) equal to the exponential of (\ref{largeLZn}), $g_1^{(n)}$ must then coincides with the free-energy density $u_n$. In turns this implies that the quantities $g_k^{(n)}$ obtained by collecting all the terms proportional to 
$(m L)^k/k!$, 
\be
g_k^{(n)} = I_{k,k} + \cdots 
\ee 
must be the $k$-power of $g_1^{(n)}$ 
\be 
g_k^{(n)} \,=\, \left[ g_1^{(n)}\right]^k \,\,\,.
\ee
Comparing the leading terms of both expression, we have   
\be
I_{k,k} \,=\,(I_{1,1})^k \,\,\,
\ee
which, written explicitly, is  
\be
\int \prod_{i=1}^k\left(\frac{d\theta_i}{2\pi} \cosh\theta_i e^{-\alpha_n q_n^{\pm}(\theta_i)}\right) |K_n(\theta_1,\ldots,\theta_n)|^2 \,=\,
\left(\int\frac{d\theta}{2\pi} \cosh\theta e^{-\alpha_n q_n^\pm(\theta)} |K_1(\theta)|^2\right)^k 
\label{finalgame}
\ee
Since, as before, this condition must hold for {\em all} charges ${\cal Q}_n$, $|K_n(\theta_1,\ldots,\theta_n)|^2$ must factorize as 
\be
|K_n(\theta_1,\ldots\,\theta_n)|^2 \,=\,\prod_{i=1}^n |K_1(\theta_i)|^2 \,\,\,.
\ee
Finally, since $\rho_d$ commutes with all conserved charges 
\be
[\rho_d,{\cal Q}_n^\pm] \,=\,0 \,\,\,,
\ee
and ${\cal Q}_n^\pm$ are a complete set of operators, $\rho_d$ is a function of them. This means that the positive function 
$|K_1(\theta)|^2$ can be written as the exponential of combination of the eigenvalues $q_n^{\pm}(\theta)$ of the conserved charges
\be
|K_1(\theta)|^2 \,=\,\exp\left[-\sum_{n=0}^\infty(\alpha_n^{+} q_n^+(\theta) + \alpha_n^{-} q_n^-(\theta))\right] \,\,\,,
\ee
where $\{\alpha_n^{\pm}\}$ are an infinite set of variables $\{\alpha_n^{\pm}\}$ which can be, in principle, fixed by 
imposing the conditions 
\be
\left.\frac{\partial {\cal Z}_n^\pm}{\partial \alpha_n^\pm}\right|_{\alpha_n^\pm =0} \,=\, L \overline{ {\cal Q}_n^{\pm}}
\,\,\,.
\ee

\newpage

\section{Infinite time averages in quench processes in the Lieb-Liniger model}

The formalism developed in the text can be applied to study the infinite-time averages of local fields in an important physical system such as the Lieb-Liniger model, a benchmark of current low-dimensional cold atom physics. The reason for that relies on the key observation, made in \cite{KMT}, that the Lieb-Liniger model, described by the Non-Linear Schr\"odinger Hamiltonian 
\be
{\cal H}\,=\,\int\mathrm{d}x\,\left(\frac{\hbar^2}{2m}\frac{\p\psid}{\p x}
\frac{\p\psi}{\p x}+\lambda\,\psid\psid\psi\psi\right)\,,
\labl{eq:HLL}
may be regarded as the non-relativistic limit of the Sinh-Gordon model. In (\ref{eq:HLL}), $\psi(x,t)$ is a 
complex Bose field $\psi(x,t)$ which satisfies the 
canonical commutation relations
\be 
[\psi(x,t),\psid(x',t)]=\delta(x-x') \,\, ,
\,\,\,\,\,\, 
[\psi(x,t),\psi(x',t)]=0 \,\,. 
\label{CRPSI}
\ee
As discussed in detail in \cite{KMT}, this mapping is realized by restoring the speed of light $c$ into the relativistic Sinh-Gordon 
Lagrangian 
\be
\mc{L}= \frac12\left(\frac{\p\phi}{c\,\p t}\right)^2-\frac12\left(\frac{\p\phi}{\p x}\right)^2-
\frac{m_0^2c^2}{g^2\hbar^2}\left(\cosh(g\,\phi)-1\right)\,.
\labl{eq:Lsg}
and taking the double limit 
\be
c\to\infty\,\,\, ,\,\,\, g\to0,\;\quad g\,c=\text{fixed}\,\,\,, 
\label{eq:limit}
\ee
where the coupling constant $\lambda$ of the Lieb-Liniger model is given by 
\be
\lambda\equiv \frac{\hbar^2c^2}{16}\,g^2\,.
\label{eq:id}
\ee
The a-dimensional coupling constant $\gamma$ of the Lieb-Liniger is $\gamma=\lambda/n$, where $n$ is the density of the one-dimensional gas. 

In this way, all physical quantities of the Sinh-Gordon model ($S$-matrix, Lagrangian and operators)
can be put in correspondence with those of the Lieb-Liniger. In particular, we can get the exact expressions of the 
matrix elements of operators such as ${\cal O}_k(x,t) = :(\psi^\dagger \psi)^k(x,t)$.

In order to show the difference which occurs between thermal equilibrium and Generalized Gibbs Ensemble averages, 
let's discuss the simplest of such operators, i.e. ${\cal O}_2(x,t)$. Its connected Form Factors are given by the 
following closed expressions 
\begin{eqnarray}
 F^{{\cal O}_2}_n(p_1,\ldots, p_n)  =  \langle p_n,\ldots , p_1 | {\cal O}_2(0,0) | p_1,\ldots, p_n \rangle 
=    \frac{1}{\lambda^3} 
\sum_{\cal P} \varphi_{LL}(p_{1,2}) \varphi_{LL}(p_{2,3}) \cdots \varphi_{LL}(p_{n-1,n}) \, p^2_{1,n} 
\end{eqnarray}
where $p_{ij}\equiv p_i-p_j$, $\sum_{{\cal P}}$ denotes the sum on all the permutations of the $\{p_j\}$ and 
\be
\varphi_{LL}(p) \,=\,\frac{2 \lambda}{p^2+\lambda^2}
\,\,\,. 
\ee
Employing now eqs. (\ref{BasicIdentity}) and (\ref{EnsembleAverage})
 for two initial states $|\alpha\rangle$ ($\alpha = a,b$) which share the same value of the energy density but not the same values of the other higher charges, and solving for them the Generalized Bethe Ansatz equation (\ref{GBA}) for the corresponding pseudo-energy, one gets the infinite time average of  $:\psi^{\dagger 2}(x)\psi^2(x): $ as function of 
the dimensionless coupling constant $\gamma$ of the Lieb-Liniger model, as shown in Fig. \ref{full}. In this figure it is also plotted, for 
comparison, the curve which corresponds to the thermal equilibrium value for this observable.

Few comments are in order. At strong coupling, i.e. $\gamma\rightarrow \infty$, all curves must go to zero for the 
emerging fermionic nature of the theory in this limit. Hence, the difference between the thermal and the 
Generalized Gibbs Ensemble averages themselves is hardly distinguishable in this region. However, at weak coupling, one 
observes a rather large spread of values for this observable, for the obvious reason that there are of course 
innumerable many ways to get out of equilibrium and the integrable theory keeps memory of it. 
In particular, one can always arrange the expectation values of the higher charges to produce a weak dependence on $\gamma$ of this observable, as it is the case for instance of the state $|b\rangle$. 

Similar computation can be repeated for other observables as well and detailed discussion will be presented in the 
forthcoming publication \cite{MDD}. 

\begin{figure}[b]
\centering
\includegraphics[width=0.65\textwidth]{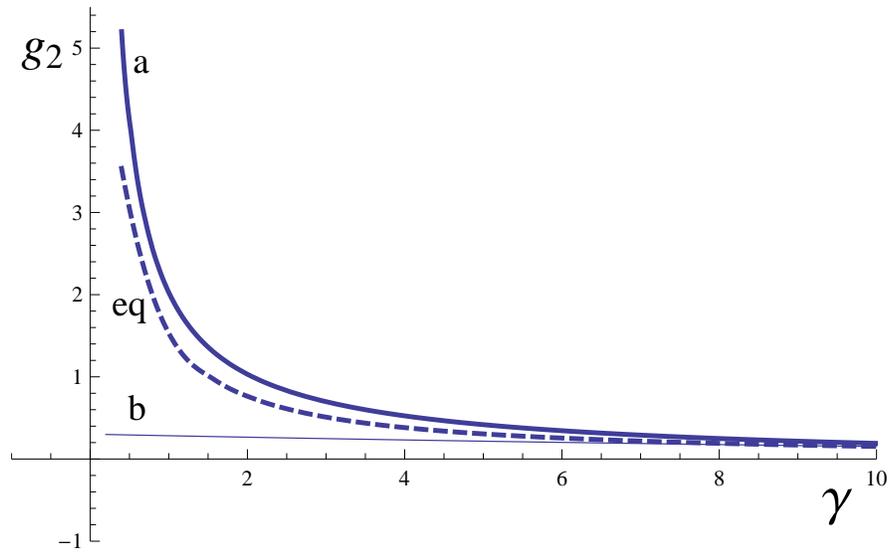} 
\caption{Infinite time averages of $\langle\alpha |(:\psi^{\dagger 2} \psi^2:) | \alpha\rangle$ ($\alpha = a,b$) for two different quench states $| a \rangle$ and $| b\rangle$ with the same energy density (here $E/L =17$) versus the dimensionless coupling constant $\gamma$ of the Lieb-Liniger model. The thermal equilibrium curve for the same energy density is the dashed one.}
\label{full}
\end{figure}


\begin{thebibliography}{XX}
\bibitem{notedisclaimer} For simplicity we consider here QIFT's with only one type of particle and with a unique ground state. The generalization to theories with many types of particles is straightforward, less trivial is the extension to theories with many vacua, a case that will be considered somewhere else.
\bibitem{note4} To appreciate the pedagogical value of this example, notice that the main differences between a free and an interacting QIFT is only at the technical level: while the free theory is solved by the Fourier Transform, an interacting QIFT needs the Inverse Scattering Transform, alias the Bethe Ansatz techniques.
\end{thebibliography}
\end{document}